\documentstyle[twoside,epsfig]{adacta}

\def\be{\begin{eqnarray}}
\def\ee{\end{eqnarray}}

\begin{document}

\prvastrana=271
\poslednastrana=280
\setcounter{page}{\prvastrana}
\headings{271}{280}

\def\autor{B.Hladk\'y et al.}
\def\nazov{Quantum synthesis of 3D vibrational states of trapped ions}

\title{\uppercase{Quantum synthesis of 3D vibrational states
of trapped ions}}

\author{B. Hladk\'y\footnote{\email{hladky@fmph.uniba.sk}}}
          {Department of Optics, Comenius University,
           Mlynsk\'a dolina, 84215 Bratislava, Slovakia}

\author{G. Drobn\'y\footnote{\email{drobny@savba.sk}},
        V. Bu\v{z}ek\footnote{also at Department of Optics,
        Comenius University, Mlynsk\'a dolina, 84215 Bratislava, Slovakia}}
        {Institute of Physics, Slovak Academy of Sciences,
         D\'ubravsk\'a cesta 9, 842 28 Bratislava, Slovakia}

\datumy{15 May 1998}{26 May 1998}

\abstract{%
A universal algorithm for a deterministic preparation
of arbitrary three--mode bosonic states is introduced.
In particular, we consider preparation of entangled quantum states
of a vibrational motion of an ion confined in a 3D trapping potential.
The target states are established after a proper sequence
of laser stimulated Raman transitions.
Stability of the algorithm with respect to a technical
noise is discussed and the distance (fidelity) of outputs with respect to
target states is studied.
}

\kapitola{1. Introduction}

The deterministic control of preparation of
genuine quantum states is one of the desired goals of quantum physics.
Recent advances in atomic physics (trapped ions \cite{Monroe1996,Cirac1996})
and quantum optics (cavity QED \cite{Brune1996})
have enabled an effective control of dynamics of microscopic quantum
systems. These systems can serve for tests of fundamental concepts of
quantum mechanics and quantum measurement theory. Other applications
can be related to quantum computing \cite{Cirac1995,Pellizzari1995}
and information processing \cite{information}.

A great effort has been devoted to the task of preparation of various
nonclassical states (e.g., Schr\"odinger cats, Fock and squeezed states)
\cite{special}.
Formal introduction of some nonclassical states (e.g., squeezed states)
has been based on the Perelomov scheme, i.e.,
under the action of an appropriate unitary operator on a reference state.
This most simple way of quantum-state preparation
can be represented as a time evolution of an input state governed by a proper
Hamiltonian. However, physical implementation of a required Hamiltonian
can face insurmountable difficulties.
Quantum-state-engineering methods introduced firstly in the cavity QED
(micromasers) have overcome these restrictions by performing conditional
measurements on atoms which result in a projection of the cavity field
on desired target states
\cite{Vogel1993,Garraway1994}.
However, due to a specific selection of quantum trajectories these methods
suffer from low probabilities of desired outputs.

The process of {\em deterministic} preparation of an arbitrary one--mode state
of electromagnetic field has been described recently by Law and Eberly
\cite{Law1996}. In their approach a proper sequence of unitary transformations
associated with a sequential switching of two interaction channels
transforms the vacuum state of the field to prescribed target wave-function.
The experimental preparation of quantum states of 1-D vibrational
motion of trapped ions has utilized switching between interaction channels
via tuning lasers either to electronic transitions or appropriate
vibrational sidebands \cite{Monroe1996}.
Possible generalizations of the 1-D quantum state synthesis
to 2-D have been discussed recently
\cite{Gardiner1997,Kneer1998,Drobny1998}.

In this paper we consider a deterministic preparation of multi--mode
bosonic states. In particular, we consider synthesis of quantum states
of a vibrational motion of an ion confined in a 3D trapping potential.
The main motivation for utilizing trapped ions is a possibility to
neglect decoherence effects due to dissipations.
An effective control of coupling between external and internal degrees of
freedom via a proper laser irradiation is also advantageous.

\kapitola{2. Quantum state synthesis}

Let us consider a preparation of an arbitrary three--mode bosonic state
\be
|\Psi_{target}\rangle =\sum_{n_x=0}^{M_x}\sum_{n_y=0}^{M_y}
                     \sum_{n_z=0}^{M_z}
                Q_{n_x,n_y,n_z} |n_x,n_y,n_z\rangle
.\label{st1}
\ee
In particular, the three modes correspond to vibrational modes of
a quantized center--of--mass motion of an ion confined in the 3D trapping
potential. The key idea of the quantum state synthesis is a splitting
of the whole Hilbert space ${\cal H}_{vib}$ of three--mode
vibrational states into a hierarchy of two--dimensional subspaces.
This enables us to transfer populations between pairs of component
Fock states (see below).
However, this dynamical ``ordering'' requires a coupling
of vibrational degrees of freedom to an adjoint system.
For the trapped ion it is natural to assume the coupling between external
(vibrational) and internal (electronic) degrees of freedom.
For purposes of our quantum-state-synthesis algorithm we utilize four
internal electronic levels $|i\rangle$ ($i=a,b,c,d$) spanning
the Hilbert space ${\cal H}_{in}$.
In general, during the preparation the vibrational and
internal states become entangled and the whole state vector can be
written as a superposition of ``component'' states
\be
|\Psi(t)\rangle &=&
\sum_{n_x=0}^{M_x}\sum_{n_y=0}^{M_y}\sum_{n_z=0}^{M_z}\sum_{i=a,b,c,d}
Q_{n_x,n_y,n_z;i}(t) |n_x,n_y,n_z\rangle \otimes |i\rangle
\label{st2} \\
&=& \sum_{J=0}^{J_{max}}\sum_{n_x=0}^{J} \sum_{n_y=0}^{J-n_x} \sum_{i=a,b,c,d}
Q_{n_x,n_y,J-n_x-n_y;i}(t) |n_x,n_y,J-n_x-n_y\rangle \otimes |i\rangle
.\nonumber
\ee
The first summation in the second expression is taken over subspaces
${\cal H}_J$ of the vibrational Hilbert space ${\cal H}_{vib}$.
The subspaces ${\cal H}_J$ are spanned with component states
$|n_x,n_y,n_z\rangle$ with a fixed total number of trap quanta
$J=n_x+n_y+n_z$.

Firstly we solve for the given target state (\ref{st1})
an inverse task, i.e. we ``de-evolve'' it towards the reference vacuum state
via a proper sequence of ``elementary'' unitary operations
which are associated with particular
laser stimulated Raman processes (see below).
The ``de-evolution'' into vacuum  can be schematically written as
(see Fig.~1)
\be
|0,0,0\rangle \otimes |a\rangle =
\left[ \hat{U}^{abc}_{0} \prod_{J=1}^{J_{max}}
\hat{U}^{ab}_{J-1,J}\hat{U}^{bcd}_{J-1}\hat{U}^{abc}_{J}
\right] |\Psi_{target}\rangle \otimes |a\rangle
\label{tr1}
.\ee
Here superscripts and subscripts indicate electronic levels and
vibrational states from subspace ${\cal H}_J$
which are involved in a controlled
``de-evolution'' of the state vector (\ref{st2}). Namely,
$\hat{U}^{abc}_{J}$ transfers populations of components states
$|n_x,n_y,n_y\rangle \otimes |i\rangle$ with $n_x+n_y+n_y=J$ and $i=a,b,c$
to a single component state  $|J,0,0\rangle \otimes |a\rangle$ which then
enters the state vector (\ref{st2}) on the subspace under consideration.
Next $\hat{U}^{bcd}_{J-1}$  {\em completely} transfers populations
of components states $|n_x,n_y,n_y\rangle \otimes |i\rangle$ with
$n_x+n_y+n_y=J-1$ and $i=b,c,d$
to the state  $|J-1,0,0\rangle \otimes |b\rangle$.
After that $\hat{U}^{ab}_{J-1,J}$ transfers the actual population of
the components state $|J,0,0\rangle \otimes |a\rangle$
to $|J-1,0,0\rangle \otimes |b\rangle$. As result, the sequence
$\hat{U}^{ab}_{J-1,J}\hat{U}^{bcd}_{J-1}\hat{U}^{abc}_{J}$ outlined in Fig.~1
cancels the contribution of component states from the Hilbert subspace
${\cal H}_J \otimes {\cal H}_{in}$ into the state vector $(\ref{st2})$.
Recursive application of this sequence on subspaces with decreasing $J$
evolves the initial target state into the vacuum.

\begin{figure}
\centerline{\epsfig{file=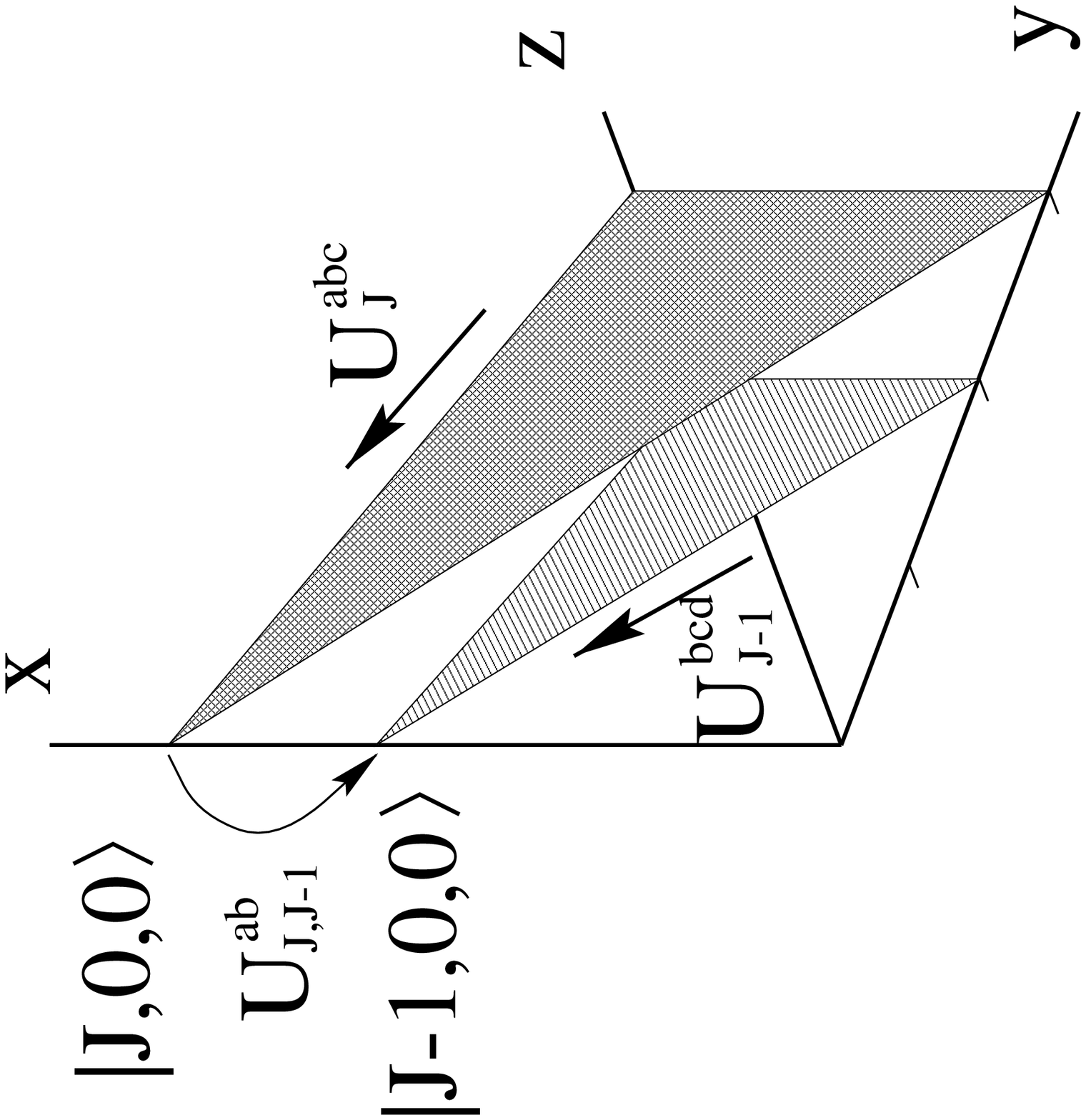,height=4.5cm,angle=-90}}
\medskip
\caption{Fig.~1. The scheme of the de-evolution procedure \protect{(\ref{tr1})}:
a recursive transfer of excitations to a subspace with a number of trap quanta
decreased by one.}
\end{figure}

Note that the solution of the inverse task (\ref{tr1})
offers immediately also the prescription for preparation of
the target state from the vacuum as
\be
|\Psi_{target}\rangle \otimes |a\rangle =
\left[  \prod_{J=J_{max}}^{1}
\hat{U}^{\dagger abc}_{J} \hat{U}^{\dagger bcd}_{J-1}
\hat{U}^{\dagger ab}_{J-1,J} \right]
\hat{U}^{\dagger abc}_{0} |0,0,0\rangle \otimes |a\rangle
\label{tr2}
.\ee
In what follows we describe in details how to ``construct''
unitary transformations appearing in Eq.(\ref{tr1})
from a set of ``elementary'' transformations (time evolution operators)
associated with laser stimulated Raman processes for a trapped ion.
Therefore the preparation sequence (\ref{tr2}) corresponds
to the action of ``elementary'' transformations applied
in the opposite order and with phases of lasers shifted
globally by $\pm\pi$.

\kapitola{3. Manipulation of trapped ions:
laser-stimulated Raman processes}

In this section we briefly introduce ``tools'' necessary
for our manipulations with vibrational states of trapped ions.
For quantum state synthesis of 3D motional states of a trapped ion
we utilize laser stimulated Raman processes \cite{Steinbach1997}.
Let us consider a trapped ion confined in a 3D harmonic potential
characterized by trap frequencies $\nu_q$ in orthogonal directions $q=x,y,z$.
The ion is irradiated along two axes ($y$ and $z$ for concreteness)
by two external laser fields with frequencies $\omega_y$, $\omega_z$
and wave vectors $k_y$, $k_z$, respectively.
The laser fields stimulate Raman transitions between two internal energy
levels $|a\rangle$ and $|b\rangle$ via an auxiliary electronic
level which is far off the resonance.
After the standard dipole and rotating--wave approximations
(RWA at laser frequencies) the adiabatic elimination of
the auxiliary off-resonant level leads to the effective interaction
Hamiltonian \cite{Steinbach1997}
\be
\hat{H}_{1}^{(eff)} =
  g^* \mbox{e}^{-i(\omega_{z}-\omega_x)t}
 \hat{D}_{y}(-i\epsilon_y) \hat{D}_{z}(i\epsilon_z) |b\rangle \langle a|
+ \mbox{h.c.}
\label{ap2}
\ee
Here $\hat{D}_{q}(i\epsilon_q)=
\mbox{e}^{i\epsilon_q(\hat{a}^{\dagger}_{q}+\hat{a}_{q})}=
\mbox{e}^{i k_q\hat{q}}$
is the displacement operator; $\hat{a}_q,\hat{a}_q^\dagger$ are
the creation and annihilation operators of the vibrational mode $q$.
The corresponding Lamb--Dicke parameter $\epsilon_q$
is defined as $\epsilon_q^2= \hbar^2 k_q^2/(2m\hbar\nu_q)$.
The effective interaction constant
$g\sim {\cal E}_y^* {\cal E}_z /\Delta$
is proportional to complex laser amplitudes ${\cal E}_q$ and
inversely proportional to detuning $\Delta$ of lasers from the auxiliary level.
The resonant terms in the Taylor series expansion of Eq.(\ref{ap2}) which
contribute dominantly to the resulting effective Hamiltonian can be selected
by an appropriate choice of laser frequencies.
If off--resonant processes are oscillating with sufficiently high frequencies
they can be eliminated applying the second RWA at trap frequencies.
In particular, tuning lasers to the first vibrational sidebands
in order to adjust resonance conditions
$\omega_1\equiv\omega_{z}-\omega_{y}=\omega_b-\omega_a+\nu_{y}-\nu_{z}$
we find that the retained resonant terms 
lead to the interaction Hamiltonian \cite{Steinbach1997}
\be
\hat{H}_1 &=&
g_1^*  \hat{a}_y^\dagger
\hat{\cal F}_{\epsilon_y}[\hat{n}_y]
\hat{\cal F}_{\epsilon_z}[\hat{n}_z] \hat{a}_z
|b\rangle \langle a| \mbox{e}^{-i\omega_1 t} + \mbox{h.c.}
,\label{ha1}
\ee
where
\be
\hat{\cal F}_{\epsilon_q}[\hat{a}_q^\dagger\hat{a}_q]&=&
\exp(-\frac{1}{2}\epsilon_q^2)
\sum_{k} \frac{(-1)^k \epsilon_q^{2k}}{(k+1)!k!}
\hat{a}_q^{\dagger k} \hat{a}_q^k
.\ee
The generalized Rabi frequency $\Omega^{(1)}_{n_y,n_z}$
of the transition between the states $|n_x,n_y,n_z\rangle\otimes |a\rangle$
and $|n_x,n_y+1,n_z-1\rangle\otimes |b\rangle$
is given as $\mbox{e}^{-\frac{\epsilon_y^2+\epsilon_z^2}{2}}
L_{n_y}^{1}(\epsilon_y^2) L_{n_z-1}^{1}(\epsilon_z^2)/\sqrt{(n_y+1)n_z}$
where $L^{1}_{n}$ is the associated Laguerre polynomial.
In the Lamb--Dicke limit of $\epsilon_q\ll 1$ it reads
$\Omega^{(1)}_{n_y,n_z}\to \sqrt{(n_y+1)n_z}$ as
$\hat{\cal F}_{\epsilon_q}$ is close to the unity operator.

With the proper laser tunings we can design also other
interaction Hamiltonians required for the quantum state
synthesis.
Namely we consider counterpropagating laser beams in the $x$ direction
which classically drive the electronic transition
$|a\rangle \leftrightarrow |b\rangle$
via a resonant laser-stimulated Raman process.
The resonance condition for laser frequencies
$\omega_2\equiv\omega_{x2}-\omega_{x1}=\omega_b-\omega_a$
leads to the interaction Hamiltonian
\be
\hat{H}_2 &=&
  g_2^* \hat{\cal F}_{\epsilon'_x}
  |b\rangle \langle a| \mbox{e}^{-i\omega_2 t} +
  g_2   \hat{\cal F}_{\epsilon'_x}
  |a\rangle \langle b| \mbox{e}^{i\omega_2 t}
\label{ha2}
.\ee
The effective Lamb-Dicke parameter
$\epsilon'_x$ with $k'=k_{x2}-k_{x1}$ is much smaller compared with
$\epsilon_x$ and thus for simplicity it is assumed to be close to zero.
Consequently $\hat{\cal F}_{\epsilon'_x}$ can be approximated by the
unity operator.
The analogous process with the laser frequencies such that
$\omega_9\equiv\omega_{x2}-\omega_{x1}=\omega_b-\omega_a-\nu_{x}$
is described by the one--mode interaction Hamiltonian $\hat{H}_3$
\be
\hat{H}_9 &=&
  g_9^* \hat{\cal F}_{\epsilon'_x}
  \hat{a}_x |b\rangle \langle a| \mbox{e}^{-i\omega_9 t} +
  g_9   \hat{\cal F}_{\epsilon'_x}
  \hat{a}_x^\dagger |a\rangle \langle b| \mbox{e}^{i\omega_9 t}
\label{ha9}
\ee
which is known as the nonlinear Jaynes-Cummings model \cite{Vogel1995}.

\kapitola{4. ``De-evolution'' procedure as sequence of
``elementary'' transformations}

Let us now assume that at some stage of the ``de-evolution'' procedure
we operate on the subspace ${\cal H}_J \otimes {\cal H}_{abc}$
and contributions  of component states from subspaces with
higher total number of trap quanta $J$ as well as
from the subspace ${\cal H}_J \otimes {\cal H}_{d}$
into the state vector $(\ref{st2})$ have been already canceled.
(The subspaces of ${\cal H}_{in}$ are spanned by indicated
electronic levels.)
Firstly, we explicitly construct the operator $\hat{U}^{abc}_{J}$
from Eq.(\ref{tr2}) which transfers {\em completely} populations of
component states from the subspace under consideration
to the single component state $|J,0,0\rangle \otimes |a\rangle$.
The operator $\hat{U}^{abc}_{J}$ can be expressed as
\be
\hat{U}^{abc}_{J}&=&\hat{A}_{J,J} \prod_{n_x=0}^{J-1}
\hat{C}_{J,n_x} \hat{B}_{J,n_x+1} \hat{A}_{J,n_x}, \qquad
\hat{C}_{J,n_x}= \hat{U}^{(5)}_{|n_x,J-n_x,0;b\rangle}
,\label{tr3}\\
\hat{A}_{J,n_x}&=&
\hat{U}^{(2)}_{|n_x,J-n_x,0;b\rangle } \hat{U}^{(1)}_{|n_x,J-1-n_x,1;a\rangle}
\ldots
\hat{U}^{(2)}_{|n_x,1,J-1-n_x;b\rangle}\hat{U}^{(1)}_{|n_x,0,J-n_x;a\rangle}
,\nonumber \\
\hat{B}_{J,n_x}&=&
\left[ \hat{U}^{(4)}_{|n_x,J-n_x,0;c\rangle } \hat{U}^{(3)}_{|n_x,J-1-n_x,1;b\rangle}
\ldots
\hat{U}^{(4)}_{|n_x,1,J-1-n_x;c\rangle} \hat{U}^{(3)}_{|n_x,0,J-n_x;b\rangle}
\right] \hat{U}^{(4)}_{|n_x,0,J-n_x;c\rangle}
\nonumber
.\ee
Here an ``elementary'' unitary transformation
$\hat{U}^{(p)}_{|n_x,n_y,n_z;i\rangle}=\exp(-i\tau\hat{H}_p)$
is equal to the evolution operator corresponding to a particular interaction
Hamiltonian $\hat{H}_p$ applied for time $\tau$.
Subscript indicates a component state $|n_x,n_y,n_z\rangle \otimes |i\rangle$
the population of which is completely transferred to another component state
according to the applied interaction ``channel'' $p$.

\begin{figure}
\begin{tabular}{cc}
\epsfig{file=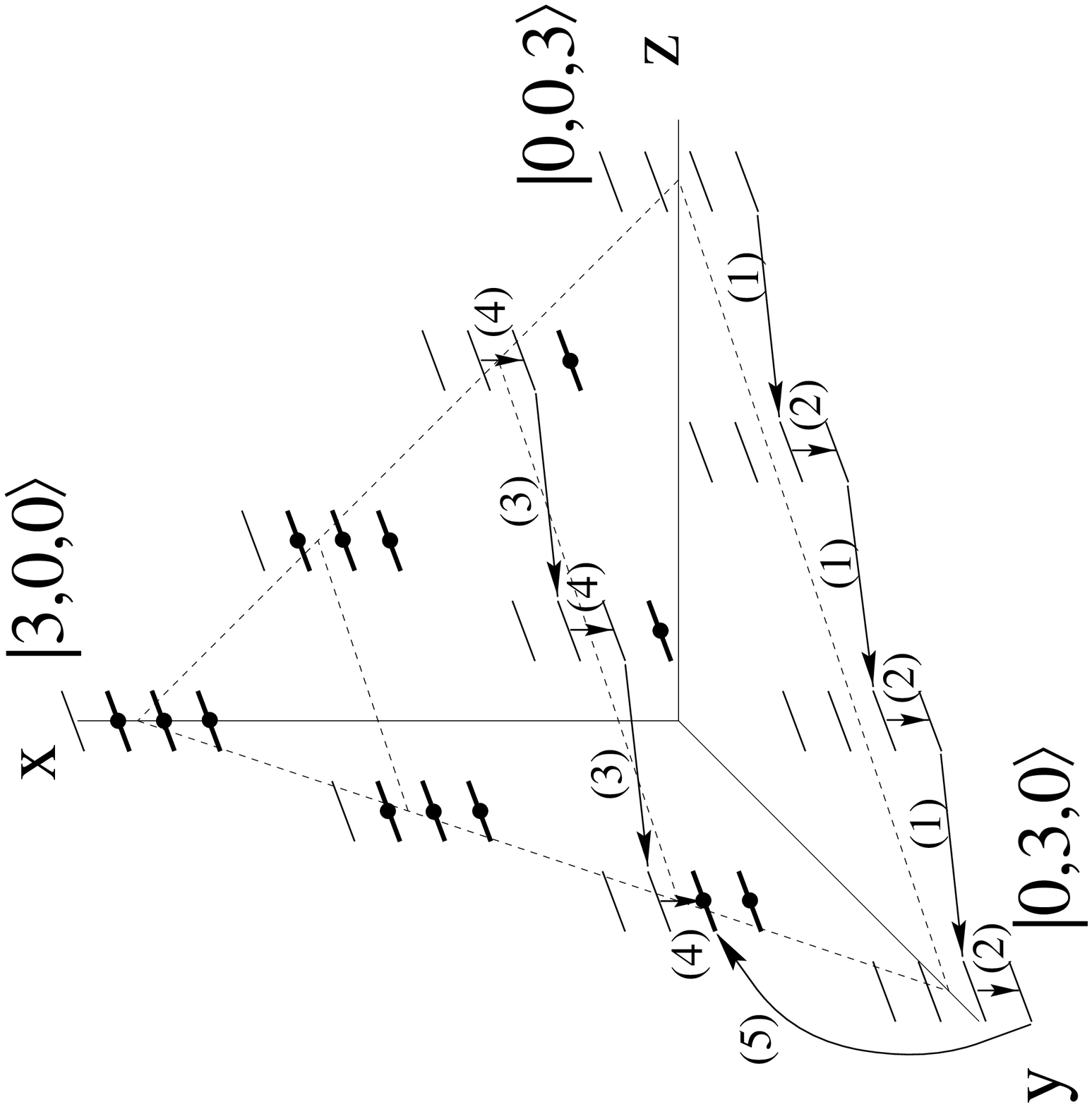,width=6.4cm,angle=-90} &
\epsfig{file=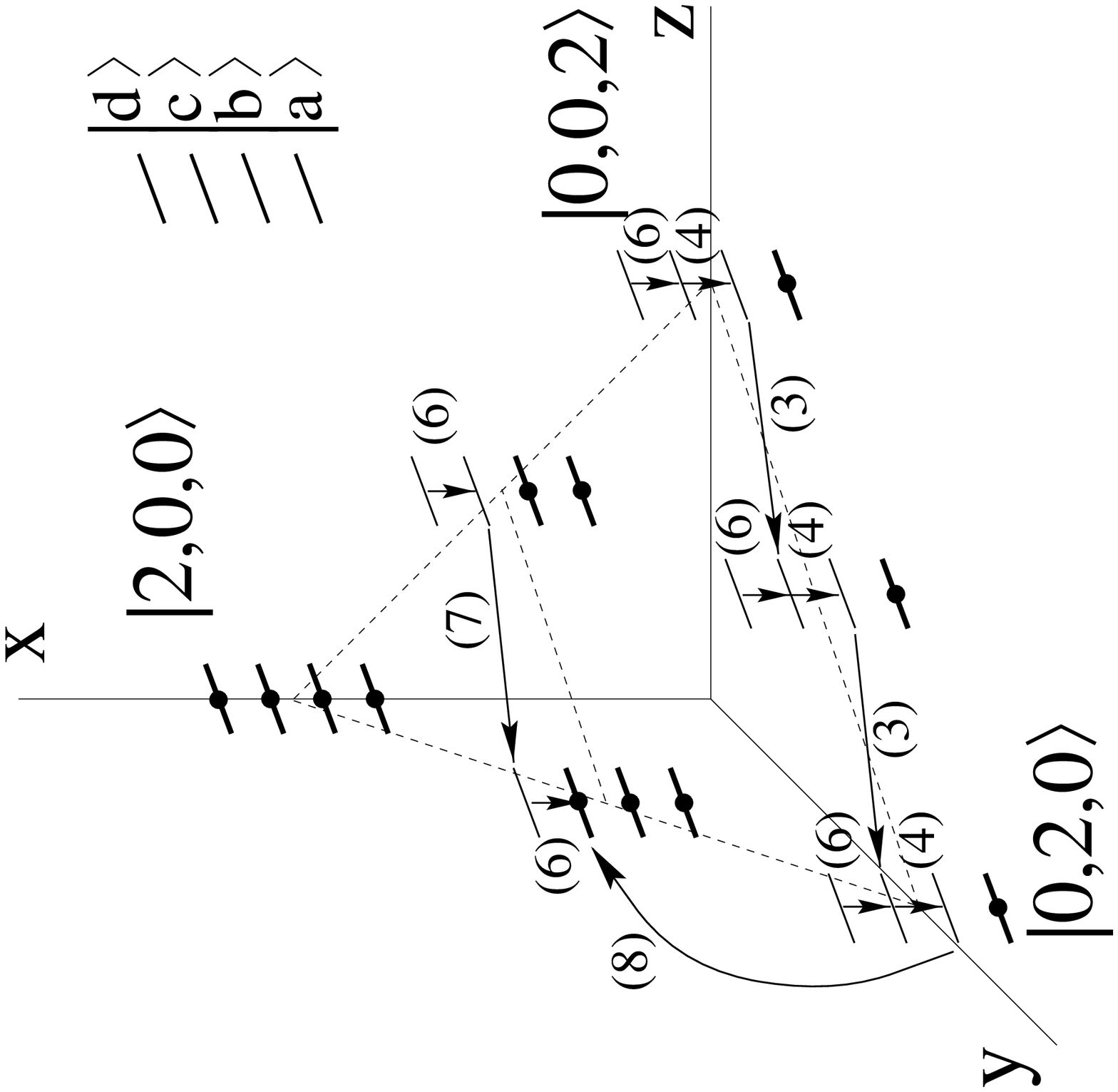,width=6.4cm,angle=-90}
\\
(a) & (b)
\end{tabular}

\medskip
\caption{Fig.~2. The scheme of the de-evolution procedure \protect{(\ref{tr1})}.
(a) Transfer of populations within the subspace with
a given number of trap quanta $J$ under action of $\hat{U}^{abc}_{J}$.
(b) The ``neighboring''subspace with the number of trap quanta
$J-1$ under action of $\hat{U}^{bcd}_{J-1}$.
Labels of arrows indicate the ``direction'' of action of the elementary 
transformations $\hat{U}^{(p)}$. 
}
\end{figure}

Actions of the ``elementary'' transformations which form
the operator $\hat{U}^{abc}_{J}$ are outlined in Fig.~2a.
In particular, under the action of
$\hat{U}^{(1)}_{|n_x,J-n_x,0;a\rangle}$ [see Eq.(\ref{tr3})]
the {\em complete} transfer of the population of the component state
$|n_x,0,J-n_x\rangle \otimes |a\rangle$
to the component state $|n_x,1,J-1-n_x\rangle \otimes |b\rangle$
is achieved providing that the interaction time $\tau$
and parameters $|g_1|,\theta_1$ of the Hamiltonian $\hat{H}_1$ (\ref{ha1})
are chosen to fulfill for $n_y=0$ the condition ($n_z=J-n_x-n_y$):
\be
\mbox{i}\mbox{e}^{-{\rm i}\theta_{1}}
Q_{n_x,n_y,n_z;a}\cos[|g_1|\tau\Omega^{(1)}_{n_y,n_z}]  +
Q_{n_x,n_y+1,n_z-1;b} \sin[|g_1|\tau\Omega^{(1)}_{n_y,n_z}] = 0
.\label{co1}
\ee
Next, the transfer of the population from  the component state
$|n_x,1,J-1-n_x\rangle \otimes |b\rangle$ to the component state
$|n_x,1,J-1-n_x\rangle \otimes |a\rangle$ is achieved under the action
of the interaction Hamiltonian $\hat{H}_2$ (\ref{ha2}).
The following choice of the interaction parameters
$|g_2| \tau$ and $\theta_2$ is required for $n_y=1$
\be
\mbox{i}\mbox{e}^{{\rm i}\theta_{2}} 
Q_{n_x,n_y,J-n_x-n_y;b}\cos(|g_2|\tau) +
Q_{n_x,n_y,J-n_x-n_y;a} \sin(|g_2|\tau) = 0
.\label{co2}
\ee
The repeated sequence of these ``elementary'' transformations forms
the operator $\hat{A}_{J,n_x}$ [see Eq.(\ref{tr3})] which transfers
the population of the states $|n_x,n_y,J-n_y-n_x\rangle \otimes |i\rangle$
with $n_y=0,\ldots,J-n_x$ and $i=a,b$ to the component state
$|n_x,J-n_x,0\rangle \otimes |a\rangle$.

Further, under the action of the operator $\hat{B}_{J,n_x+1}$
populations of states $|n_x+1,n_y,J-n_y-(n_x+1)\rangle \otimes |i\rangle$
with $n_y=0,\ldots,J-(n_x+1)$ and $i=b,c$ are transferred to the component
state $|n_x+1,J-(n_x+1),0\rangle \otimes |b\rangle$.
It is seen from Eq.(\ref{tr3}) that the form of $\hat{B}_{J,n_x+1}$
(acting on ``neighboring lines'' in Fig.~2a with number of $x$-quanta
increased by one) is quite analogous to $\hat{A}_{J,n_x}$
but instead of levels $|a\rangle,|b\rangle$
the electronic levels $|b\rangle,|c\rangle$
are involved in the transfer of populations. The corresponding Hamiltonians
$\hat{H}_3$, $\hat{H}_4$ and the transfer conditions are obtained
from Eqs.(\ref{ha1}),(\ref{co1}) and (\ref{ha2}),(\ref{co2})
by the substitutions: $|a\rangle\to |b\rangle$ and $|b\rangle\to |c\rangle$.

\noindent 
As the next step of the ``de-evolution'' sequence (\ref{tr3}) the operator
$\hat{C}_{J,n_x}= \hat{U}^{(5)}_{|n_x,J-n_x,0;b\rangle}$ is applied
to the state vector (\ref{st2}). It transfers the whole population
of $|n_x,J-n_x,0\rangle \otimes |a\rangle$ to
$|n_x+1,J-(n_x+1),0\rangle \otimes |b\rangle$.
The corresponding interaction Hamiltonian $\hat{H}_5$ is obtained from
$\hat{H}_1$ after replacements $y\to x$, $z\to y$. In other words,
two lasers stimulating the Raman transition operate in $x,y$ directions
instead of $y,z$. Note that the action of the operator $\hat{B}_{J,n_x+1}$
was necessary to avoid a transfer of population backwards
to component states with smaller $n_x$ on the given subspace
${\cal H}_J\otimes {\cal H}_{abc}$.

Altogether, the r\^{o}le of the operator $\hat{U}_J^{abc}$ (\ref{tr3})
is to transfer populations of all component states from
${\cal H}_J\otimes {\cal H}_{abc}$ to
the only component state $|J,0,0\rangle \otimes |a\rangle$.
The next important step in the ``de-evolution'' procedure (\ref{tr1})
is related to the operator $\hat{U}_{J-1}^{bcd}$ which
transfers populations of all component states
from the ``neighboring'' subspace
${\cal H}_{J-1}\otimes {\cal H}_{bcd}$ to
its only component state $|J-1,0,0\rangle \otimes |b\rangle$.
The operator $\hat{U}_{J}^{bcd}$ can be built up from a set of ``elementary''
transformations in a close analogy with the unitary transformation
$\hat{U}_{J}^{abc}$ by utilizing the electronic levels $i=b,c,d$
instead of $i=a,b,c$. The action of $\hat{U}_{J-1}^{bcd}$
represented by a sequence of ``elementary'' transformations
is outlined in Fig.~2b.
The interaction Hamiltonians $\hat{H}_6$ and $\hat{H}_7$ associated
with the indicated ``elementary'' transformations are obtained
from $\hat{H}_4$ and $\hat{H}_3$, respectively,
by the substitutions $|c\rangle\to |d\rangle$ and $|b\rangle\to |c\rangle$.
The Hamiltonian $\hat{H}_8$ arises from $\hat{H}_5$
after substitutions $|b\rangle\to |c\rangle$
and $|a\rangle\to |b\rangle$.

The next step which connects the subspaces shown in Figs.~2a,b
has been already sketched in Fig.~1. The operator
$\hat{U}^{ab}_{J-1,J}=\hat{U}^{(9)}_{|J,0,0;a\rangle}$
associated with the interaction Hamiltonian (\ref{ha9})
transfers the population of the component state
$|J,0,0\rangle \otimes |a\rangle$ to $|J-1,0,0\rangle \otimes |b\rangle$.
In other words, the operator $\hat{U}^{ab}_{J-1,J}$ moves us from
the subspace ${\cal H}_J \otimes {\cal H}_{abc}$
to the subspace ${\cal H}_{J-1} \otimes {\cal H}_{abc}$.
Note that contribution from ${\cal H}_{J-1} \otimes {\cal H}_{d}$
into the state vector (\ref{st2}) has been canceled by the previous action
of $\hat{U}^{bcd}_{J-1}$.
It means that we can apply the whole ``de-evolution''
procedure (\ref{tr1}) recursively decreasing the total number
of trap quanta.

\begin{figure}[t]
\vspace*{-0.5cm}
\centerline{\epsfig{file=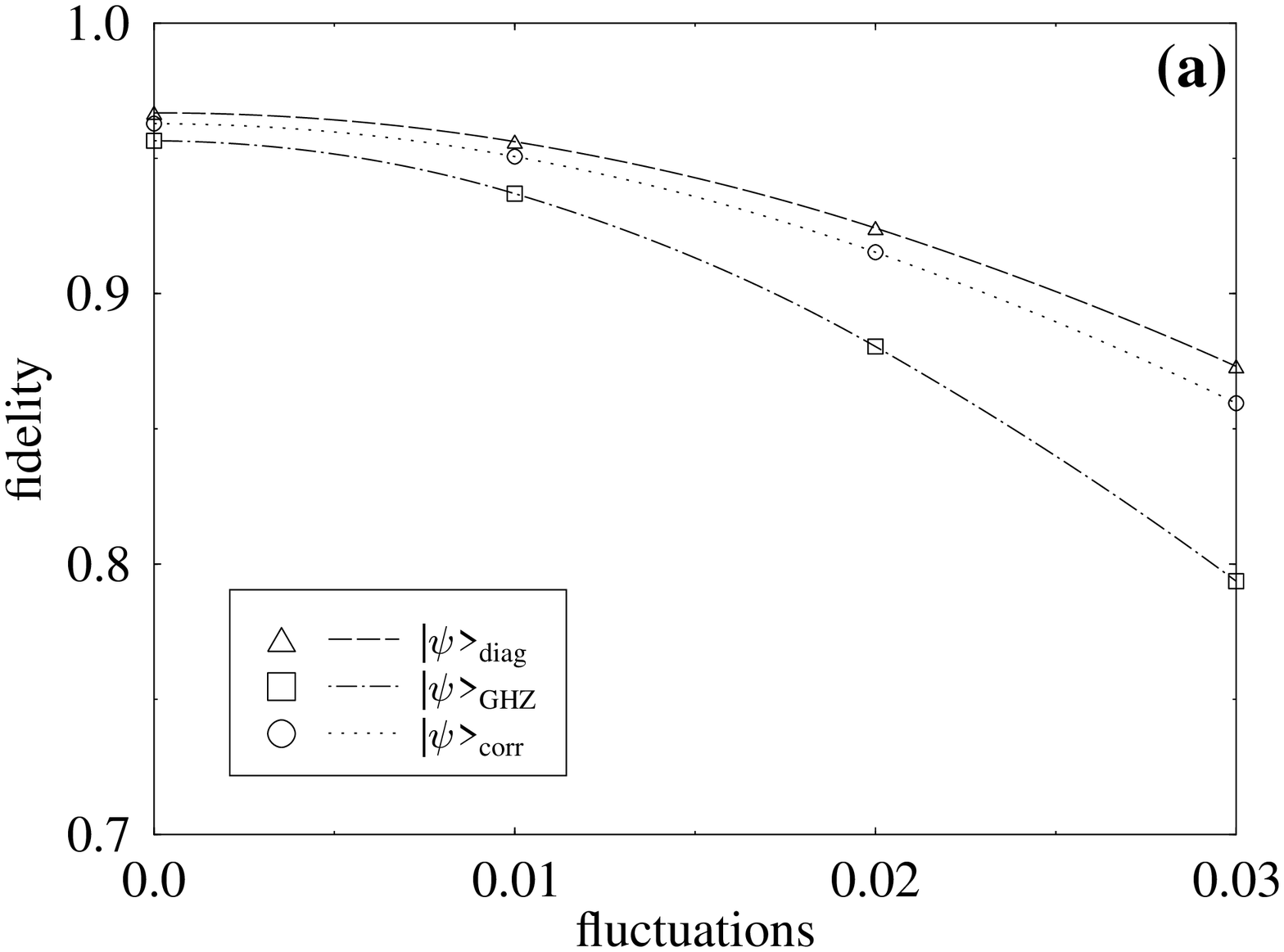,width=8cm}}
\centerline{\epsfig{file=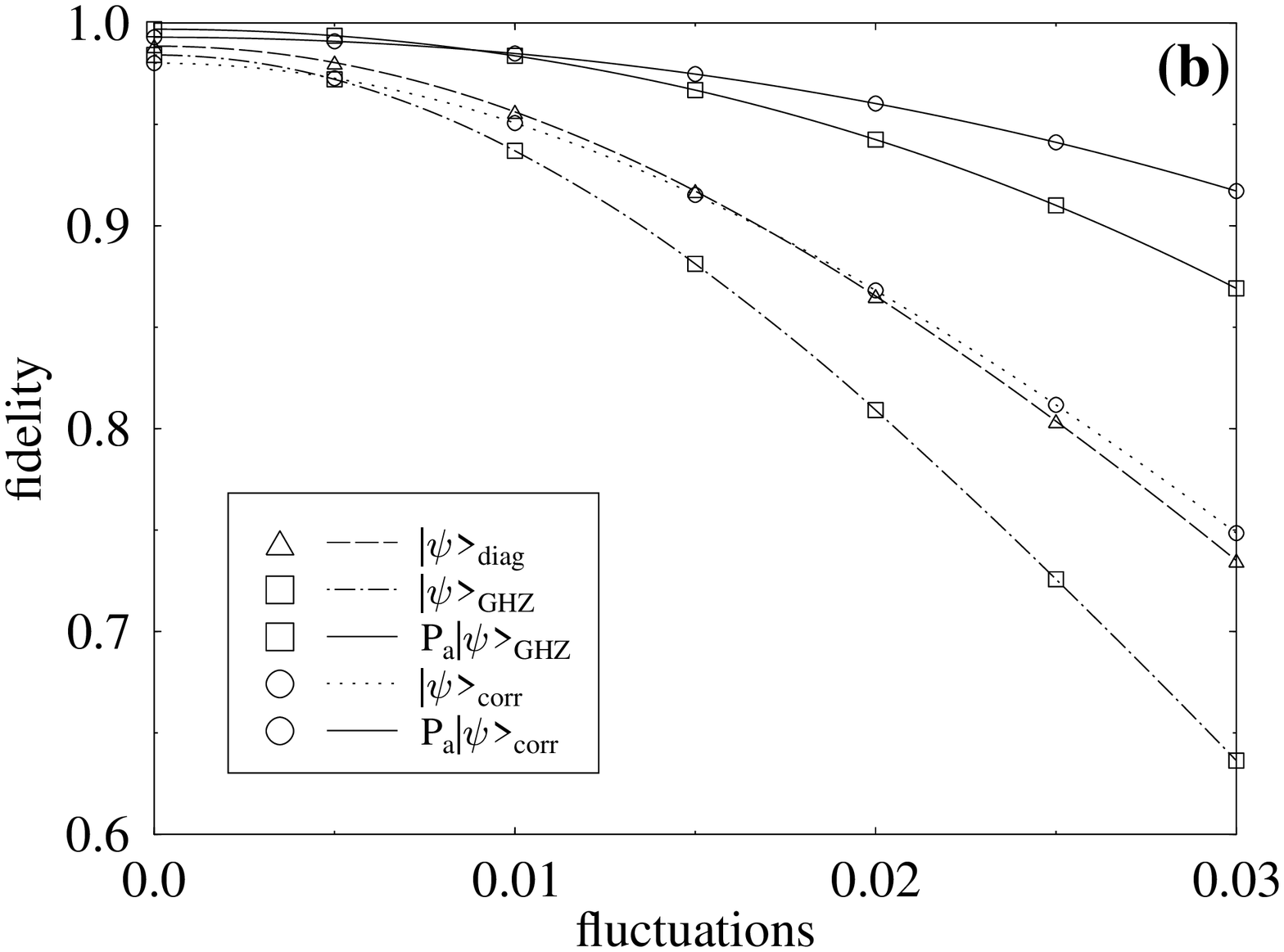,width=8cm}}

\caption{Fig.~3. Fidelities of the outputs to the indicated target states
in the presence of a technical noise. 
(a) Varying phase fluctuations $\delta_{\theta}$ 
with fixed fluctuations $\delta=0.01$ of the interaction lengths.
(b) Varying fluctuations $\delta$ of interaction lengths 
with fixed phase fluctuations $\delta_{\theta}=0.01$.
}
\end{figure}

\kapitola{5. Remarks on imperfect preparation}

The {\em deterministic} preparation procedure as described above is based
on the unitary time evolution (\ref{tr2}) of the reference state
to the target state. The solution for the inverse problem
(\ref{tr1}) can be found only in the absence of dissipations.
However, one important source of noise (even for almost dissipation--free
dynamics of trapped ions) is an imperfect control
of interaction constants (which depend on laser amplitudes and phases)
and switching times required for transfer of populations of component
states.
To estimate the influence of this ``technical noise''
now we introduce random fluctuations
around the ideal interaction constants and switching times defined by
(\ref{tr2}). In particular, we assume that fluctuations are
equally distributed around the ideal values of interaction lengths 
$|g_p|\tau$ [e.g., solutions of Eqs.(\ref{co1},\ref{co2})]
within a given interval $\delta$. Also phases fluctuate
around their ideal values $\theta_p$ within a fixed interval
$\delta_{\theta}$.
The preparation sequence (\ref{tr2}) with non-ideal
interaction constants and switching times then results in some
output state $|\Psi_{\delta}\rangle$ of the form (\ref{st2}).
Its internal and external degrees of freedom are in general entangled.
The deviation from the desired target state
$|\Psi_{target}\rangle\otimes |a\rangle$ can be quantified
using an average fidelity of particular imperfect realizations
$|\Psi_\delta\rangle$ to the target state:
\be
f_{\delta}=\langle\!\langle |\langle\Psi_{\delta}|
\Psi_{target}\rangle|a\rangle|^2 \rangle\!\rangle_{\delta}
,\label{fid}
\ee
where $\langle\!\langle \ldots \rangle\!\rangle_{\delta}$
stands for the average
over imperfect realizations with a given range of fluctuations
$\delta$.
We have performed simulations of noisy quantum state synthesis
for three specific target states. We have considered
the three-mode ``Greenberger--Horne--Zeilinger'' state
$|\Psi_{GHZ}\rangle={\cal N}
(|\alpha,\alpha,\alpha\rangle + |-\alpha,-\alpha,-\alpha\rangle)$
with coherent components; three--mode correlated state
$|\Psi_{corr}\rangle=
{\rm e}^{-|\alpha|^2/2} \sum_{n} \frac{\alpha^n}{\sqrt{n!}} |n,n,n\rangle$,
with $\alpha =1$; and the superposition state
$|\Psi_{diag}\rangle=
\frac{1}{\sqrt{5}} \sum_{n=0}^{4} |n,n,n\rangle$.
The cutoff in Fock basis is chosen as $J_{max}=10$.
The corresponding average fidelities (over 100 noisy preparations)
are shown in Fig.~3a for an increasing interval of phase fluctuations 
$\delta_{\theta}$ around the ideal phases $\theta_p$.
The interaction lengths fluctuate around ideal values 
$|g_p|\tau$ within a fixed interval $\delta=0.01$. 
The Lamb--Dicke parameters are chosen as 
$\epsilon_x=0.3$, $\epsilon_y=0.1$, $\epsilon_z=0.2$, $\epsilon'_x=0.1$.
In Fig.~3b we show averaged fidelities for increasing fluctuations $\delta$ 
of the interaction lengths, now with fixed phase fluctuations 
$\delta_{\theta}=0.01$.

The figures clearly demonstrate that the preparation procedure works well
even in the presence of relatively small fluctuations.
In general, the overall error is accumulated by each ``elementary''
transformation. The total number of ``elementary'' transformations scales
polynomially with respect to the dimensionality of the Hilbert space.
In our particular case the total number of operations is proportional
to $J_{max}^3$. We have simulated the noisy preparation procedure 
also for higher values of the Lamb--Dicke parameters (up to
$\epsilon_q=0.4$). The dependence of fidelities on amplitudes of fluctuations
is very similar to that one shown in Fig.~3.
In particular, it turns out that the fidelities are less sensitive
to (absolute) fluctuations $\delta$ of the interaction lengths when increasing 
the Lamb--Dicke parameters. This is due to an increase of the ideal
values $|g_p|\tau$ of the interactions lengths required for 
the quantum-state synthesis.

One possibility to improve fidelities of outputs is to utilize
conditional measurements of the internal electronic levels.
The preparation procedure starts and ends with the ion in the
same internal state $|a\rangle$. So after preparation sequence we can
verify whether the ion occupies the desired internal state.
It can be done by driving dipole transitions from other internal levels
to proper auxiliary levels and observing
fluorescence signals. No signal means that the undisturbed ion is in
the right electronic state $|a\rangle$.
Wrong preparation sequences (with the test signal) are thrown away.
In other words, outputs are projected on $\hat{P}_a=|a\rangle \langle a|$.
The improvement of fidelities due to the conditional selection of outputs
is shown in Fig.~3b.
Corresponding decrease of probabilities to find on the check
the right final internal state is not so dramatic and state-dependent
as known from conditional methods for micromasers.
For example, the fidelity is lifted from 75\% to 92\% (64\% $\to$ 87\%)
for the state $|\Psi_{corr}\rangle$ ($|\Psi_{GHZ}\rangle$)
with fluctuations $\delta=0.03$ with the efficiency 82\% (74\%).

\kapitola{6. Summary}

We have investigated preparation of entangled quantum states
of the vibrational motion of the ion confined in the 3D trapping potential.
We introduced the universal algorithm for deterministic preparation
of quantum states
which is based on a sequence of laser stimulated Raman transitions.
In the presence of a technical noise fidelities of outputs
with respect to desired target states can be significantly improved
when conditional measurements on the electronic levels of the ion
are performed.
The method can be adopted also for other three--mode bosonic systems.

\medskip
\noindent {\bf Acknowledgement}
We thank Jason Twamley for helpful discussions.


\end{document}